\def\vday{{\bf 8/6/98}}
\documentstyle[12pt]{article}
\def\baselinestretch{1.5}
\begin{document}
\thispagestyle{empty}
\null\vskip -1cm
\centerline{
\vbox{
\hbox{August 1998}\vskip -9pt
\hbox{\em\bf\vday}\vskip -9pt
\hbox{hep-ph/9808252}\vskip -9pt
     }
\hfill
\vbox{
\hbox{UICHEP-TH/98-8}
     }     } \vskip 1cm

\centerline{\large \bf    CP Violation in the Width Difference}
\centerline{\large \bf    of the Scalar Electron Decay due}
\centerline{\large \bf    to the Complex Gaugino Masses}
\vspace{1cm}
\centerline           {
Fred Browning$^{(1)}$,
Darwin Chang$^{(2)}$,
and
Wai-Yee Keung$^{(1,3)}$ }
\begin{center}
\it
$^{(1)}$Physics Department, University of Illinois at Chicago,
IL 60607-7059, USA\\
$^{(2)}$NCTS and Physics Department, National Tsing Hua University, Hsinchu, 
Taiwan\\ 
$^{(3)}$KEK, 1-1 Oho, Tsukuba-shi, Ibaraki-ken 305, Japan
\vspace{.5cm}
\end{center}

\begin{abstract}
We calculate the width difference of the scalar electron decay,
$ \Gamma(\tilde e_L         \rightarrow      e_L\omega)
- \Gamma({\bar{\tilde e}}_L \rightarrow \bar e_L\omega)$ in the softly
broken supersymmetrized standard model. The CP asymmetry is assumed to 
arise from the complex gaugino masses.  Even in limit that the electron 
mass $m_e$ is vanishing, the difference can be of the order of $\alpha$.
\end{abstract}
\vspace{1in}


\newpage

\section*{Introduction}
Supersymmetric extensions of the Standard Model (SM) has been widely 
considered to be the most likely new physics beyond SM.  In particular, 
such extension can contains various new mechanisms of CP violation.  It 
is the purpose of this note to explore the consequence one of such mechanism.

   In the softly broken supersymmetrized standard model\cite{reviews}, the
dimension-3 gaugino masses can be complex, {\it i.e.}
$-{1\over2}m_\psi \ {\rm e}^{-i2\phi}
\overline{(\psi_L)^c}\psi_L +{\rm h.c.}$, 
which can be transformed into the canonical Majorana form
$-{1\over2} m_\psi\bar\psi\psi$ with the Majorana field $\psi$  defined to be
${\rm e}^{-i\phi} \psi_L+\ {\rm e}^{i\phi}(\psi_L)^c$. 
However, such rearrangement only shift the absorbed phase into interaction 
terms such as gauge interactions. %
For example, since the electron has the quantum numbers
\begin{equation}
\begin{array} {|c|c|c|} \hline
{}~ & Y/2        &    T_3     \\
\hline
e_L & -{1\over2} & -{1\over2} \\
\hline
e_R & -1         &     0      \\
\hline
\end{array} \quad ,
\end{equation}
its supersymmetric couplings to the scalar electron and the gauginos
[the bino $\beta$ and the ({\it neutral}) wino $\omega$] become complex, and 
can be tabulated as:
\begin{equation}
\begin{array} {|c|c|c|} \hline
{}~ & {\rm Bino\ } \beta  & {\rm Wino\ } \omega  \\
\hline
e_L & {\sqrt{2} e\over 2 \cos\theta_W }\ {\rm e}^{i\phi_\beta} 
    & {\sqrt{2} e\over 2 \sin\theta_W }\ {\rm e}^{i\phi_\omega} \\
\hline
e_R & {\sqrt{2} e\over   \cos\theta_W }\ {\rm e}^{i\phi_\beta} 
    & 0 \\
\hline
\end{array} \quad .
\end{equation}
The supersymmetric part of gauge interaction Lagrangian is
\begin{equation}
{\cal L}=
 {\sqrt{2} e\over 2 \cos\theta_W }\ {\rm e}^{i\phi_\beta}
\tilde e_L \bar e_L \beta
+
 {\sqrt{2} e\over 2 \sin\theta_W }\ {\rm e}^{i\phi_\omega}
\tilde e_L \bar e_L \omega
+
 {\sqrt{2} e\over    \cos\theta_W }\ {\rm e}^{i\phi_\beta}
\tilde e_R \bar e_R \beta  +{\rm \ h.c.} + \cdots \ .
\end{equation}
The CP phase dependence of the CP-even parameters in the cross section of 
$e_L e_L \rightarrow \tilde e_L \tilde e_L$ has been recently studied
in Ref.\cite{scott}. However, it is important also to establish direct CP
violation by searching  for CP-odd effects, such as  decay width
differences. In this article, we calculate the asymmetry between conjugated
channels, like 
$ \Gamma(\tilde e_L         \rightarrow      e_L\omega)
-\Gamma({\bar{\tilde e}}_L \rightarrow \bar e_L\omega)$.
It is clear that in the limit when the electron mass is vanishing, only 
the phase difference $\phi_\omega-\phi_\beta$ is physical in the above 
Lagrangian.

\section*{Formalism}
Without loss of generality, we assume the bino $\beta$ and the wino
$\omega$ do not have significant mixing with the Higgsinos.
To simplify our discussions, we also assume that $\tilde{e}_L$ and 
$\tilde{e}_R$ do not mix.  If they do mix, it will constitute another 
mechanism  of CP violation which should be treated separately. 
In this limit, electric dipole measurements are not directly sensitive
to the phase difference $\phi_\omega-\phi_\beta$, which can be of order
unity.
It is very likely that both these weak gauginos, $\beta$ and $\omega$, are
lighter than the scalar left handed electron $\tilde e_L$. In this case
 the asymmetry mentioned above exists through the final state
interaction between the two decay channels. 
We denote physical masses of $\tilde e_L$, $\beta$, and $\omega$
as  $m$, $m_1$, and $m_2$, respectively.  Note that $\tilde{e}_R$ is 
never involved in our process.
The amplitude at the tree level for the process
$\tilde e_L(P) \rightarrow e_L(p) \omega(p')$ is

\begin{equation}
{\cal M}^0=
\left({\sqrt{2} e {\rm \ e}^{i\phi_\omega} \over 2 \sin\theta_W}\right) 
\bar u(e_L,p){1+\gamma^5\over 2}v(\omega,p')
\quad .
\end{equation}

The one-loop amplitude occurs with the intermediate state $e_L\beta$.
The Feynman diagram, as shown in Fig.~1, for  the process,
$\tilde e_L(P)\rightarrow \beta(k) e_L(k')\rightarrow
e_L(p)\omega(p')$ gives

\begin{eqnarray}
{\cal M}^1 &=& \left( {\sqrt{2} e {\rm \  e}^{-i\phi_\omega}
                                    \over 2 \sin\theta_W}\right) 
           \left(i{\sqrt{2} e  {\rm \  e}^{i\phi_\beta}
                                    \over 2  \cos\theta_W}\right)^2
              {id^4q/(2\pi)^4\over q^2-m^2}     \nonumber\\
&&\  \times \ 
\bar u(e_L,p) {1+\gamma^5\over2} {i(\not k+m_1)\over k^2-m_1^2}
              {1+\gamma^5\over2} {(-i\not k')\over {k'}^2}
              {1-\gamma^5\over2}  v(\omega,p')
\ .
\end{eqnarray}
The amplitude involves the integration of the virtual momentum $q$
which runs through the loop.  
We use the Feynman rules adopted in 
Refs.\cite{majorana-keung}
to deal with
couplings of Majorana fermions. 
The rules were summarized in Ref.\cite{majorana-denner}.
Since the dispersive part of the amplitude will not produce CP
violation in the decay width difference at the 1-loop level, in the
following we keep only the absorptive part which is easily 
obtained\cite{cut-rule}
from the above expression by requiring the on-shell condition
$k^2=m_1^2$ and ${k'}^2=0$, removing the corresponding denominators, and
integrating over the solid angle $\Omega$ of $k$ in the rest frame of
$\tilde e_L$. This method of calculation is very similar to those of
CP violation in baryogensis\cite{barr}  or in Higgs decays\cite{ref:dcH}.
\begin{equation}
{\cal M}^1 ={i\over 16\pi}
 {\sqrt{2} e {\rm \ e}^{-i\phi_\omega} 
\over 2 \sin\theta_W}
           {e^2 {\rm \ e}^{i2\phi_\beta}
\over 2 \cos^2\theta_W}
\bar u(e_L,p) {1+\gamma^5\over2} m_1 (\not q+\not p') v(\omega,p')
              {1-m_1^2/s\over q^2-m^2}{d\Omega\over4\pi}
\ .
\end{equation}
Although the incoming momentum squared $s$ and the scalar electron
mass squared $m^2$ in the $t$-channel propagator are the same in our
case, we use different symbols for them above for possible extension of the 
formulas. %
First, we need to establish the following  covariant expression,
\begin{equation}
\int {q^\mu\over q^2-m^2} {d\Omega\over 4\pi}=Ap^\mu+B{p'}^\mu
\ .
\end{equation}
Only  the $B$ term is relevant to our calculation. It is obtained by the
dot products with $p$ on both sides of the above equation.
\begin{equation}
B={1\over s-m_2^2}
   \left( 1+{(m^2-m_1^2)s\over(s-m_2^2)(s-m_1^2)}
\ln{m^2-m_1^2m_2^2/s\over s+m^2-m_1^2-m_2^2} \right)
\ .
\end{equation}
Finally we obtain,
\begin{equation}
{\cal M}^1=-\left( {\sqrt{2} e {\rm \ e}^{-i\phi_\omega}
\over 2 \sin\theta_W}\right) 
           \left({i\alpha {\rm \ e}^{i2\phi_\beta}
\over 8 \cos^2\theta_W}\right)
\bar u(e_L,p) {1+\gamma^5\over2}  v(\omega,p')
{m_1m_2 \over s} {s-m_1^2\over s-m_2^2}\ \times \ {\cal I}
\ ,
\end{equation}
where
\begin{equation}
{\cal I}=1+{s+m^2-m_1^2-m_2^2\over (s-m_1^2)(s-m_2^2)}s
\ln{m^2-m_1^2m_2^2/s\over s+m^2-m_1^2-m_2^2}
\ .
\end{equation}
This finally gives
\begin{equation}
{\cal A}^\omega={
 \Gamma(\tilde e_L         \rightarrow      e_L\omega)
-\Gamma({\bar{\tilde e}}_L \rightarrow \bar e_L\omega)
\over
{1\over2}
[ \Gamma(\tilde e_L         \rightarrow      e_L\omega)
+\Gamma(\bar{\tilde{e}}_L  \rightarrow \bar e_L\omega)
]}
=-{ \alpha \sin(2\phi)\over 2\cos^2\theta_W} {m_1m_2\over s}
{s-m_1^2\over s-m_2^2} \ \times \ {\cal I}
\ .
\end{equation}
with $\phi=\phi_\omega-\phi_\beta$, {\it i.e.} the CP violation here
depends only on the relative phase.  Note that when either $m_1$ or $m_2$ 
goes to zero, the asymmetry disappears as expected
because the phase $\phi$ loses its meaning in the limit $m_e = 0$.
The asymmetry is of the order $\alpha$, and there is no $m_e$
suppression. Similarily, the complementary asymmetry is
\begin{equation}
{\cal A}^\beta={
 \Gamma(\tilde e_L         \rightarrow      e_L\beta)
-\Gamma({\bar{\tilde e}}_L \rightarrow \bar e_L\beta)
\over
{1\over2}
[ \Gamma(\tilde e_L         \rightarrow      e_L\beta)
+\Gamma(\bar{\tilde{e}}_L  \rightarrow \bar e_L\beta)
]}
={ \alpha \sin(2\phi)\over 2\sin^2\theta_W} {m_1m_2\over s}
{s-m_2^2\over s-m_1^2} \ \times \ {\cal I}
\ .
\end{equation}
These expressions respect the CPT relations,
\begin{equation}
 \Gamma(\tilde e_L         \rightarrow      e_L\omega)
-\Gamma({\bar{\tilde e}}_L \rightarrow \bar e_L\omega)
=-\Gamma(\tilde e_L         \rightarrow      e_L\beta)
+\Gamma({\bar{\tilde e}}_L \rightarrow \bar e_L\beta) \quad .
\end{equation}

\section*{Phenomenology}

In $e^+e^-$ annihilation at high enough energy, scalar leptons
can be copiously produced\cite{peskin} in
pairs. If the bino is the lightest supersymmetric particle, the event
profiles of the two channels $\tilde{e}_L\to e_L \beta$ and 
$\tilde{e}_L\to e_L
\omega$ are very different. In the $e_L\beta$ mode, the bino $\beta$ just
carries away the missing momentum quietly.  In the $e_L\omega$ mode,
the wino $\omega$ continues to decay, $\omega \to e_L{\tilde e}_L^*
\to e_L {\bar e}_L \beta$, producing $e^+e^-$ pair and missing momentum. 
Distinguishing these two kinds of decay channels, we can test their CP
asymmetry in the decay branching fractions. In this article, we have
shown that such CP asymmetry can be of the order of $\alpha$. 
It is not large,
see Fig. 2, but {\it not} suppressed by the tiny factor $m_e^2/m^2$.

\section*{Acknowledgement}

F.~B. and W.-Y.~K. are supported by a grant from the Department of
Energy, and D.~C. by a grant from the National Science Council of
R.O.C.
W.-Y.~K. thanks the hospitality and support of the KEK theory group,
and useful discussions with V. Barger and T. Falk on gaugino phases.

\newpage

\section*{\bf Figure Caption}

\begin{itemize}
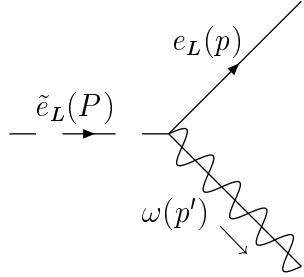
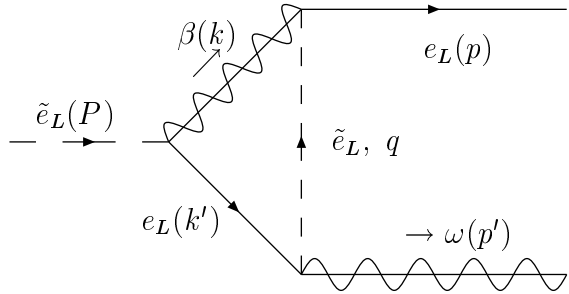
\item[Fig. 1.]  
(a) Tree level diagrams for $\tilde e_L\rightarrow e_L\omega$;
(b) One-loop diagram for $\tilde e_L\rightarrow e_L\omega$.
\item[Fig. 2.]
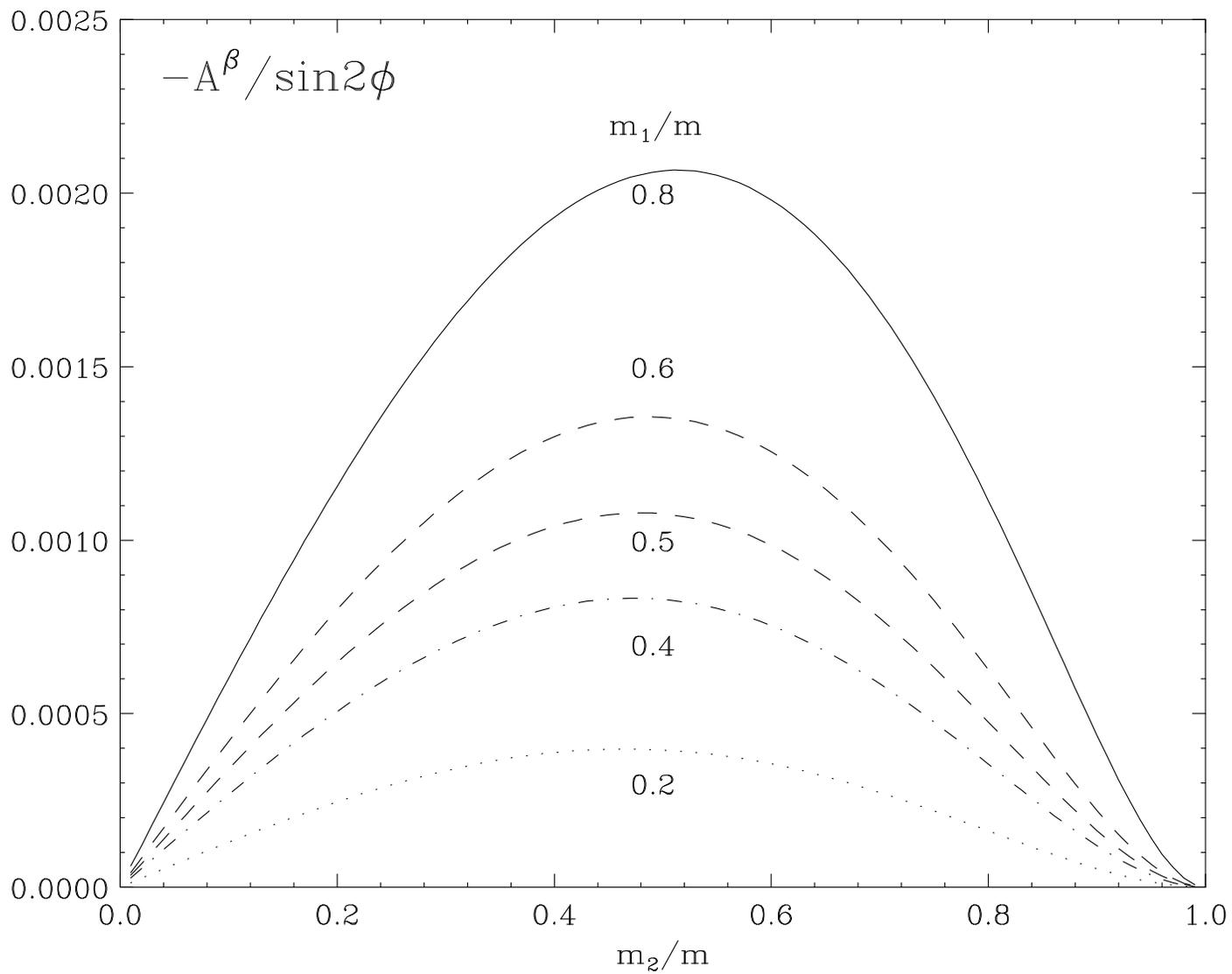
Plot of asymmetry $A^\beta/\sin(2\phi)$.

\end{itemize}

\def\baselinestretch{1.2}

\vfill\eject

\end{document}